\documentclass[11pt, executivepaper]{article}
\usepackage[utf8]{inputenc}
\usepackage[T1]{fontenc}
\usepackage{natbib}
\usepackage{amsmath}
\usepackage{mathtools}
\usepackage{xcolor}
\usepackage{amsfonts}
\usepackage{graphicx}
\usepackage{enumitem}
\usepackage{geometry}
\usepackage{hyperref}
\hypersetup{colorlinks= true, allcolors=blue}
\setcitestyle{aysep={}}
\begin{document}

\title{\textbf{On the Classification between $\psi-$Ontic and $\psi-$Epistemic Ontological Models}\thanks{The authors contributed equally to this paper.}}

\author{Andrea Oldofredi\thanks{Contact Information: University of Lausanne, Dept.\ of Philosophy,
1015 Lausanne, Switzerland. E-mail: Andrea.Oldofredi@unil.ch} \and Cristian Lopez\thanks{University of Lausanne, Dept.\ of Philosophy,
1015 Lausanne, Switzerland and University of Buenos Aires, Argentina. E-mail: lopez.cristian1987@gmail.com.}}

\maketitle

\begin{center}
\emph{Accepted for publication in Foundations of Physics}
\end{center}
\vspace{5mm}

\begin{abstract}
\cite{Harrigan:2010} provided a categorization of quantum ontological models classifying them as $\psi-$ontic or $\psi-$epistemic if the quantum state $\psi$ describes respectively either a physical reality or mere observers' knowledge. Moreover, they claimed that Einstein---who was a supporter of the statistical interpretation of quantum mechanics---endorsed an epistemic view of $\psi$. In this essay we critically assess such a classification and some of its consequences by proposing a two-fold argumentation. Firstly, we show that Harrigan and Spekkens' categorization implicitly assumes that a complete description of a quantum system (its ontic state, $\lambda$) only concerns \emph{single, individual} systems instantiating \emph{absolute, intrinsic} properties. Secondly, we argue that such assumptions conflict with some current interpretations of quantum mechanics, which employ different ontic states as a complete description of quantum systems. In particular, we will show that, since in the statistical interpretation ontic states describe ensembles rather than individuals, such a view  cannot be considered $\psi-$epistemic. As a consequence, the authors misinterpreted Einstein's view concerning the nature of the quantum state. Next, we will focus on \emph{Relational Quantum Mechanics} and \emph{Perspectival Quantum Mechanics}, which in virtue of their relational and perspectival metaphysics employ ontic states $\lambda$ dealing with relational properties. We conclude that Harrigan and Spekkens' categorization is too narrow and entails an inadequate classification of the mentioned interpretations of quantum theory. Hence, any satisfactory classification of quantum ontological models ought to take into account the variations of $\lambda$ across different interpretations of quantum mechanics.
\vspace{5mm}

\end{abstract}
\clearpage 

\tableofcontents
\vspace{10mm}

\section{Introduction}
\label{Intro}

The question concerning what is the correct interpretation of Quantum Mechanics (QM) and how it represents the microphysical world arose since the first formulations of the theory in the twenties. The fervent debates about the status and meaning of the quantum state---which are still alive today---clearly witness this fact. Against this background, there is some agreement among experts working on the foundations of quantum theory that, if such a framework tells us in some relevant sense what the world is like, then the ontological status of the quantum state $\psi$ must be clearly spelled out.\ Thus far, different interpretations of the quantum formalism faced in various ways the metaphysical issue of what is the nature of the quantum state, and how it relates to the underlying reality (if it does it at all). Consequently, it is not surprising that $\psi$ has been given different statuses and roles.

Referring to this, the first crucial question in this inquiry is likely to be the following: Does the quantum state represent a state of some physical entity out there in the world, or does it represent instead a state of our knowledge about something out there in the world? Interestingly, answers to this question roughly outline two stances regarding the nature of the quantum state: if it is assumed to represent real physical systems, it is said to be \emph{ontic}; otherwise, it is considered \emph{epistemic}. Hence, one can sort interpretations out depending on whether the quantum state is considered as representing some state of affairs in the world, or just agents' knowledge. In addition, when discussing the interpretation of the quantum state, one has to take into account another distinction as crucial as the former. Indeed, $\psi$ may represent the state of a physical quantum system either \emph{completely} or \emph{partially}; thus, interpretations of QM can be sorted out depending on whether they consider the quantum state to be a complete or a partial description of physical systems. In the latter case, a complete description of the state of a system may require some supplementary set of variables---as exemplified by the class of hidden variables models and theories.

Although these different views about (and categorizations of) the quantum state were somehow known since the very beginning of quantum theory, a rigorous classification has been given only recently in \cite{Harrigan:2010}, where the authors provided strict criteria to define the ontic or epistemic nature of quantum states. Even though one of the main aims of Harrigan and Spekkens is to provide a novel classification for \emph{hidden variable} models, they clearly go beyond this scope, since their classification also allows accommodating non-hidden variable models, extending the categories to almost any interpretation of QM.\footnote{In this essay we assume that readers have some familiarity with Harrigan and Spekkens' work.} For this reason, Harrigan and Spekkens' categorization of models have obtained a far-reaching resonance, and soon became common currency among physicists and philosophers of physics when discussing not only hidden variable models, but also the meaning of $\psi$ in various interpretations of QM (cf.\ for instance \cite{Gao:2017} and \cite{Cabello:2017}). Nonetheless, although Harrigan and Spekkens' distinction has significantly contributed to shed clarity on a complex debate, in this paper we will argue that there are some relevant aspects that their classification either overlooks or fails to fully capture. 

More precisely, we will discuss some assumptions underlying Harrigan and Spekkens' classification, with particular attention to (i) their definition of  ``complete physical state'' of quantum systems represented by their ontic state $\lambda$---and, thereby, the structure of the ontic space $\Lambda$---and (ii) the relations between $\lambda$ and $\psi$. Indeed, the nature of $\lambda$, as defined by these authors, has been so far scarcely discussed in the literature, which has rather focused on how $\psi$ represents $\lambda$. This fact, unfortunately, has led to some misrepresentations of the nature of the ontic state in relation to various interpretations of QM. Harrigan and Spekkens' classification, in fact, implicitly assumes (i) that unalike ontological models pose the very same $\lambda$, and (ii) that they just disagree about the relation held between $\lambda$ and $\psi$. Clearly, there would be many ways to unfold different assumptions with respect to $\lambda$, however, in the present essay we will be concerned with the following. The authors assume that, in any ontological model

\begin{enumerate}
\item $\lambda$ represents the state of a \emph{single, individual quantum system};
\item $\lambda$ is \emph{perspective independent}.
\end{enumerate}

We will argue that both assumptions are unwarranted and might be a source of confusion when assessing some interpretations of QM. Firstly, it is worth noting that one of the principal aims of the authors is to argue that Einstein not only showed that quantum theory is incomplete, but also that the quantum state represents merely observers' knowledge, consequently endorsing a $\psi-$epistemic view. Given that he supported a statistical view of QM, Harrigan and Spekkens conclude that such an interpretation must be considered $\psi-$epistemic. However, not every quantum theory presupposes that $\lambda$ represents the state of a \emph{single, individual} quantum system. Paradigmatically, the \emph{statistical} (or \emph{ensemble}) interpretation of quantum mechanics (cf.\ \cite{Ballentine:1970}) assumes that $\psi$ provides a description of the statistical properties of an ensemble of similarly prepared systems. In this case, however, there is no matter of fact to suppose that a statistical ontological model is also committed to single, individual systems that it fails to describe. On the contrary, an accurate reading of such a view suggests that statistical ontological models do not take $\lambda$ as providing a complete description of a single, individual quantum system, but of an ensemble. That is, a statistical ontological model poses a different sort of $\lambda$ with respect to that employed in \cite{Harrigan:2010}. This fact, in turn, is conceptually relevant since it implies that the statistical interpretation should not be considered a $\psi-$incomplete and $\psi-$epistemic model---as in Harrigan and Spekkens' classification---but $\psi-$complete and, thereby, $\psi-$ontic. We will also discuss the implications of this fact with respect to Harrigan and Spekkens' interpretation of Einstein's view about the quantum state.

Secondly, we will claim that not every ontological model should assume that $\lambda$ is perspective independent. This is crucial for some \emph{relational} and \emph{perspectival} interpretations of QM\footnote{In the remainder of the essay we will sometime refer to relational and perspectival interpretations of QM as Relational and Perspectival \emph{Quantum Mechanics}, whereas we will frequently refer to Ballentine's or Einstein's views as the ensemble or statistical \emph{interpretation} of QM. However, we do not mean any conceptual difference between them: we take all these as different interpretations of the standard formalism and not as alternative formulations of QM.}, where it would be meaningless to claim that $\lambda$ is independent of a given reference system or perspective (cf.\ \cite{Rovelli:1996} and \cite{Dieks:2019} respectively). Remarkably, one of the consequences of not taking the relational nature of $\lambda$ into account is to conflate relationalism and/or perspectivalism with subjectivism and incompleteness with respect to the quantum state. On the contrary, being perspectival or relational should not be regarded exclusively as a property of $\psi$, but also---and more importantly---as a property of $\lambda$. In the particular case of Rovelli's interpretation, although we agree with Harrigan and Spekkens in saying that the quantum state is $\psi-$epistemic, we do not consider such an interpretation to be $\psi-$incomplete: even in the absence of an absolute perspective to describe $\lambda$ quantum mechanically, when the tenets of Rovelli's theory are seriously taken into account, it is not fair to consider $\psi$ as incomplete. Moreover, in other forms of quantum perspectivalism, for instance that defended by Dieks, the quantum state is evidently $\psi-$ontic, though different perspectives could deliver different $\psi$s for a perspectival $\lambda$.

The structure of the paper is the following: in Section \ref{2} we briefly explain Harrigan and Spekkens' classification. In Section \ref{3}, we will show that $\lambda$ must \emph{not} be always regarded as the state of a single, individual system. To exemplify this point we take into consideration the ensemble interpretation of QM. In Section \ref{4}, we aim at showing that $\lambda$ should not be regarded as perspective independent either. In this case, we will consider Rovelli's relational and Dieks' perspectival interpretations of QM to support our theses. Finally, conclusions are drawn in Section \ref{5}.

\section{$\psi-$Ontic and $\psi-$Epistemic Ontological Models}
\label{2}

\cite{Harrigan:2010} provide a stratified classification of ontological quantum models that essentially depends on the nature of the quantum state. Such a classification is suitable for the authors' purpose, since they aim primarily to show that Einstein's interpretation of QM---which was statistical in essence (cf.\ \cite{Einstein:1936} and \cite{Einstein:1949})---supported a $\psi-$epistemic view of the quantum state. More precisely, Harrigan and Spekkens argue not only that Einstein showed that the quantum mechanical wave function is not a complete representation of physical systems, but also, and more importantly, that $\psi$ is epistemic, i.e.\ it describes not an underlying physical reality, but merely observers' knowledge. Since this thesis will be critically assessed in the reminder of the present essay, let us introduce in more detail Harrigan and Spekkens classification starting from the definitions of what ontological models are within their approach. 

It is worth noting that the authors define ontological models employing an operational setting, i.e.\ the primitive notions of such models consist exclusively in preparations procedures of physical systems in certain states and measurements performed on them. A complete specification of the properties of a given physical system is provided by $\lambda$, the ontic state of the system under scrutiny. Referring to this ``in an operational formulation of quantum theory, every preparation $P$ is associated with a density operator $\rho$ on Hilbert space, and every measurement $M$ is associated with a positive operator valued measure (POVM) $\{ E_k\}$. (In special cases, these may be associated with vectors in Hilbert space and Hermitian operators respectively.) The probability of obtaining outcome $k$ is given by the generalized Born rule, $p(k|M, P)=\textrm{Tr}(\rho E_k)$'' (\cite{Harrigan:2010}, p.\ 128). Indeed, the main aim of such operational models is to provide probabilities $p(k|M, P)$ of outcomes $k$ for some measurement $M$ performed on the prepared systems, given the set of preparation instructions $P$.\ When a measurement is carried out, a measuring device will ``reveal something about those properties'' (\emph{ibid.}). 
To this regard, Harrigan and Spekkens  underline that, although an agent may know the preparation procedures prior the performance of a certain measurement, she may have incomplete knowledge of $\lambda$. Thus, it follows that while outcomes $k$ are determined by $\lambda$, and therefore the probability to obtain them is $p(k|\lambda, M)$, the epistemic state of the observer is given by $p(\lambda| P)$---notably, an observer which has incomplete information about $\lambda$ assigns ``non-sharp'' probability distributions over the ontic space $\Lambda$. Hence, the authors do not specify the metaphysical nature of quantum objects at play. In this context, the word ``ontological'' is used in a weaker sense with respect to the standard philosophical jargon: the authors simply mean that the quantum models just refer to something real in the world, leaving however unspecified the details concerning their ontologies.

Let us now introduce the first categorization of models.\footnote{In this section we do not follow exactly Harrigan and Spekkens' order in presenting the several categorizations of their approach, since the authors introduce in the first place the distinction between $\psi-$ontic and $\psi-$epistemic models. However, for ease of exposition, we prefer to introduce in the first place the distinction between $\psi-$complete and $\psi-$incomplete models.}\ Harrigan and Spekkens divide quantum theories in $\psi-$\emph{complete} and $\psi-$\emph{incomplete}: in the former case, the quantum state encodes every information about systems which refer to real physical objects existing in the world. Alternatively stated, in $\psi$-complete views the quantum state represents completely the features of a physical system. Indeed, it is generally accepted that, according to standard QM, $\psi$ provides such a complete description of quantum systems and, consequently, it is taken to describe an underlying reality; for this reason \cite{Harrigan:2010} defined this theoretical framework as $\psi-$complete. Contrary to standard QM, hidden variables models supply $\psi$ with additional variables in order to represent the states of physical systems; thus, such models are defined as $\psi-$\emph{incomplete} insofar as the quantum state provides only a partial description of the features of physical systems. Remarkably, to the extent to which $\psi$ represents only partially the state of the quantum system and needs to be supplemented with additional structures, hidden variable models are also called $\psi-$supplemented by Harrigan and Spekkens.

In the second place, the authors propose another distinction concerning the status of quantum states, classifying $\psi$ either as \emph{ontic} or \emph{epistemic}. In a nutshell, a model is defined $\psi-$ontic if every complete physical state (i.e.\ the ontic state $\lambda$) can be consistently described by a pure state. According to Harrigan and Spekkens' categorization, in $\psi-$ontic models different quantum states correspond to disjoint probability distributions over the space of ontic states $\Lambda$. More precisely, a model is said to be $\psi-$ontic if ``for any pair of preparation procedures, $P_{\psi}$ and $P_{\phi}$, associated with distinct quantum states $\psi$ and $\phi$, we have $p(\lambda | P_{\psi})p(\lambda|P_{\phi})=0$ for all $\lambda$'' (\cite{Harrigan:2010}, pp.\ 131-132), meaning that observers' epistemic states associated with different quantum states are non-overlapping. On the contrary, a model is defined $\psi-$epistemic if there exist ontic states consistent with more than one pure state. Therefore, in such epistemic models, there are quantum states that correspond to overlapping probability distributions in $\Lambda$.
Indeed, this is a distinctive feature of these models since it implies that agents' epistemic states may overlap, i.e.\ there exist preparation procedures $P_{\psi}, P_{\phi}$ such that $p(\lambda | P_{\psi})p(\lambda|P_{\phi})\neq0$, meaning that the ontic state $\lambda$ can be consistently represented by both quantum states $\psi$ and $\phi$: ``[i]n a $\psi-$epistemic model, multiple distinct quantum states are consistent with the same state of reality---the ontic state $\lambda$ does not encode $\psi$'' (\emph{ibid.}, p.\ 132). This explains why the authors claim that in the case of $\psi-$epistemic models, the quantum state refers to observers' incomplete knowledge of reality, and not to reality itself.

Now we are ready to combine both distinctions in order to classify different ontological models. Considering $\psi-$complete models, there is a one-to-one relation between reality and its complete description provided by the pure quantum state $\psi$. Consequently, in these models, knowing the quantum state implies having a complete knowledge of the ontic state of the system under consideration. Therefore, $\psi-$complete models are necessarily also $\psi-$ontic. Examples of such models are given by standard QM, Everett's relative-state formulation (cf.\ \cite{Everett:1957aa}), the Many-World interpretation (cf.\ \cite{Wallace:2012aa}), and Wave-Function Realism (cf.\ \cite{Albert:2013}). 

If ontological models are $\psi-$incomplete, then they may be either $\psi-$supplemented or $\psi-$epistemic. In the former case, the description of a physical system is supplemented by some additional (or hidden) variables, whose value is generally unknown. In hidden variables models, trivially, the quantum state provides partially or incomplete knowledge of the system. Examples of hidden variables models are given by Bohmian mechanics (cf.\ \cite{Durr:2013aa}), Bohm's pilot-wave theory (cf.\ \cite{Bohm:1952aa}), and Nelson's stochastic mechanics (cf.\ \cite{Nelson:1966aa}). Notably, also the class of $\psi-$supplemented models is $\psi-$ontic. Finally, in the case of $\psi-$epistemic models, $\psi$ represents an agent's incomplete knowledge of reality, and not reality itself. A typical example of such a kind of models is given by QBism (cf.\ \cite{Fuchs:2002}).

From these distinctions some conclusions can be inferred. In the first place, models which are simultaneously $\psi-$complete \emph{and} $\psi-$epistemic cannot exist. Therefore, if a model is $\psi-$complete, it must be $\psi-$ontic (cf.\ Lemma 6, \cite{Harrigan:2010}, p.\ 133). Alternatively, if a model is $\psi-$incomplete, then it can either be $\psi-$ontic, as in the case of hidden variable models, or $\psi-$epistemic. If a model is $\psi-$epistemic, then it cannot in any case be $\psi-$ontic, since it does not describe any underlying physical reality, but only the agents' knowledge of it. 

In the remainder of this essay we will critically analyze the premises and presuppositions of Harrigan and Spekkens' distinctions. We will explain why the thesis according to which Einstein's interpretation of QM is  $\psi-$epistemic can be hardly defended. Moreover, we will argue that some assumptions concerning the nature of $\lambda$ not only seem to be unwarranted, but also conflict some tenets of current interpretations of QM. 

\section{The Statistical Interpretation of QM: $\psi-$ontic or $\psi-$epistemic?}
\label{3}

One of the main claims of Harrigan and Spekkens is that Einstein endorsed a $\psi$-epistemic view of the quantum state. The authors primarily support this claim by analyzing one of Einstein's arguments aimed at demonstrating the incompleteness of quantum mechanics contained in a letter to Schr\"odinger dated June 19th, 1935. Given this claim and that Einstein was a supporter of the statistical view of the quantum state, Harrigan and Spekkens conclude that such an interpretation must be considered $\psi-$epistemic.\footnote{Also \cite{Leifer:2014} put the statistical interpretation in the $\psi-$epistemic camp.} In this section we will argue for the following theses: (i) to maintain that Einstein held a $\psi-$epistemic view misrepresents his thoughts on QM, and (ii) that it is not correct to conceive the statistical interpretation of QM as a $\psi-$epistemic model (at least without making further assumptions). Hence, we here underline an important limitation of Harrigan and Spekkens' classification, which can also have negative consequences for their account of Einstein's position with respect to the interpretation of QM. 

\subsection{Einstein and the $\psi-$epistemic view}

Let us explain in the first place the reasons for which the categorization introduced in the previous section classifies the ensemble view as $\psi-$epistemic. As we have seen, ontological models consider operational procedures to prepare the state of a quantum system in a certain manner as primitive notions. Such procedures are associated with some observable properties, whose values will be then revealed by the performance of a set of measurements on the physical system under scrutiny. The ontic state $\lambda$ of such a system is supposed to specify all its properties, i.e.\ it is supposed to describe it completely. Furthermore, the ontic space $\Lambda$ encodes the reality envisaged by the model of the system under consideration in a very special sense: in Harrigan and Spekkens' account, $\lambda$ represents exclusively individual systems, not ensemble of systems. If there is a one-to-one relation between $\psi$ and $\lambda$, then, the description provided by $\psi$ is complete, i.e.\ it provides a complete description of an individual system.\ However, to affirm that $\psi$ does not yield a complete description of a quantum system---as stated in several places by Einstein\footnote{Einstein famously expressed his view about the incompleteness of quantum mechanics in the essay written with Boris Podolsky and Nathan Rosen resulting in the well-known EPR paradox (\cite{Einstein:1935}). However, it is a notorious historical fact that Einstein was dissatisfied with the EPR paper, which had been written by Podolsky. Einstein then reformulated his own incompleteness argument in many other places, as for instance in his correspondence with Schr\"odinger---where he introduced the though experiment now widely known as ``Einstein's boxes'' and published in \cite{Einstein:1936} (see also \cite{Norsen:2005} for a nice discussion)---in his correspondence with Max Born, in the essay ``Quanten-Mechanik und Wirklichkeit'' published in \emph{Dialectica} in 1948, and in his intellectual autobiography contained in \cite{Einstein:1949}.}---is to claim that $\psi$ must be incomplete as long as it does not match with $\lambda$. Thus, according to Harrigan and Spekkens' classification, $\psi$ must be either $\psi$-supplemented, or $\psi$-epistemic. Insomuch as the statistical view does not introduce or propose hidden variables, this interpretation cannot be considered a $\psi-$supplemented model. Therefore, as a simple logical consequence of the authors' classification, the ensemble view must be a $\psi-$incomplete model which is also $\psi-$epistemic. Consequently, the quantum state is taken to represent the  observer's knowledge of the system's state.

It is now worth stressing the importance of such a conclusion for Harrigan and Spekkens' hypothesis concerning Einstein's support of a $\psi-$epistemic view. In the first place, the authors carefully discuss Einstein's attempt aimed at showing the incompleteness of the quantum mechanical description of physical systems. However, they do not take into consideration the EPR paradox, but rather, a simplified and clearer argument contained in a letter to Schr\"odinger dated June 19th, 1935. In this correspondence Einstein stated that, in his opinion, QM would have been a complete theory only if the quantum state $\psi$ would be correlated one-to-one with ``the real state of the real system''. A failure of this requirement would then naturally imply the incompleteness of the theory:
\begin{quote}
In the quantum theory, one describes a real state of a system through a normalized function, $\psi$, of the coordinates (of the configuration-space) [...] Now one would like to say the following: $\psi$ is correlated one-to-one with the real state of the real system. [...] If this works, then I speak of a complete description of reality by the theory. But if such an interpretation is not feasible, I call the theoretical description `incomplete' (Einstein's letter to Schr\"odinger, June 19th 1935, quote in \cite{Howard:1985}, p.\ 179).
\end{quote}

\noindent Referring to this, Harrigan and Spekkens establish a direct connection between Einstein's criterion of completeness and their own definition of $\psi-$completeness, which reads:
\begin{quote}
An ontological model is $\psi-$\emph{complete} if the ontic state space $\Lambda$ is isomorphic to the projective Hilbert space $\mathcal{PH}$ (the space of rays of Hilbert space) and if every preparation procedure $P_{\psi}$ associated in quantum theory with a given ray $\psi$ is associated in the ontological model with a Dirac delta function centered at the ontic state $\lambda_{\psi}$ that is isomorphic to $\psi$, $p(\lambda|P_{\psi})=\delta(\lambda-\lambda_{\psi})$ (\cite{Harrigan:2010}, p.\ 131).
\end{quote} 

\noindent In other words, according to them the ``real state of the real system'' is precisely the parameter $\lambda$ referring to the ontic state of a certain system. Regarding this, it should be underlined that Einstein's criterion of completeness and Harrigan and Spekkens' definition are grounded on different bases: on the one hand, the latter authors start from an operational perspective, for which $\lambda$ describes properties associated with the preparation procedures chosen by an observer in a given experimental situation.\ On the other hand, Einstein had in mind a strong ontological reading of the completeness criterion: QM should include every physical feature of systems in their quantum states, where these properties do refer to the inherent attributes instantiated by the system.\ As also stressed by \cite{Howard:1985}, Einstein thought that systems do possess definite attributes at all times, independently of measurements, being such a belief firmly grounded in the separability principle. Nonetheless, for the sake of the discussion, this important difference can be left aside. Let us then assume that these definitions are pragmatically equivalent in the precise sense that, for both Einstein and Harrigan and Spekkens, in order for QM to be complete, there must be a one-to-one relation between $\psi$ and the real state of physical systems, that is, $\lambda$.

Einstein's argument proceeds as follows. Firstly, one has to take into consideration, as in the EPR scenario, an entangled pair formed by two systems $A$ and $B$ which, after their interaction, form the joint system $AB$. Secondly, the subsystems $A$ and $B$ are sent to space-like separated regions. Finally, Einstein notes that the performance of a measurement of a two-valued operator upon $A$ will affect the result in $B$, where the quantum state may have values $\psi_B, \psi_{\underline{B}}$. Hence: 

\begin{quote}
what is essential is exclusively that $\psi_B$ and $\psi_{\underline{B}}$ are in general different from one another. I assert that this difference is incompatible with the hypothesis that the description is correlated one-to-one with the physical reality (the real state). After the collision, the real state of ($AB$) consists precisely of the real state of $A$ and the real state of $B$, which two states have nothing to do with one another. The real state of $B$ thus cannot depend upon the kind of measurement I carry out on $A$. (Separation hypothesis from above.) But then for the same state of $B$ there are two (in general arbitrarily many) equally justified $\psi_B$ which contradicts the hypothesis of a one-to-one or complete description of the real states (Einstein's letter to Schr\"odinger, June 19th 1935, quoted in \cite{Harrigan:2010}, p.\ 147).
\end{quote} 

As Howard underlines, this proof of incompleteness differs from that included in the EPR paper, being only indirect and involving a contradiction between his criterion of completeness and the separability principle. The latter principle entails that $B$'s real state should remain unaffected by manipulation, observation and measurement performed on the space-like separated system $A$. Hence, the different quantum states $\psi_B, \psi_{\underline{B}}$ attributable to $B$ on the basis of the different observation on $A$ must be correlated with the same real state of $B$. Indeed, as Howard writes ``if the $\psi-$function were to provide a complete description of the real state of $B$, it would have to be correlated one-to-one with $B$'s real state, and there we have our contradiction with the separation principle'' (\cite{Howard:1985}, p.\ 180).
In sum, Einstein's criterion of completeness employed in this argument implies that different quantum states must be correlated with different real, ontic states of the systems. Here is the key for Harrigan and Spekkens crucial conclusion: since there exist two possible states that one may assign to $B$ as results of measurements on the system $A$, there exist two possible epistemic states in overlap since both of them are ascribable to $B$. Therefore, based on their own definitions, the authors conclude not only that $\psi$ fails to be complete---as Einstein's argument shows---but also that it fails to be ontic! According to them, Einstein affirms that there may be multiple quantum states associated with the same ontic state, i.e.\ for the state $B$ there would be two equally possible and equally justified states $\psi_{B}$ and $\psi_{\underline{B}}$. To state it differently, changes in $\psi$ would not correspond to changes in the ontic state. Therefore, Harrigan and Spekkens conclude that Einstein showed the failure of $\psi-$onticity; consequently, $\psi-$complete and $\psi-$supplemented models cannot represent his views on QM. Thus, the authors claim, one is left with just one option: Einstein sought to adopt a $\psi-$incomplete and $\psi-$epistemic interpretation of the quantum state. 

\subsection{Einstein, the Statistical Interpretation of QM and $\psi-$completeness}
 
Although Harrigan and Spekkens' interpretation of Einstein's view may seem correct at first sight, in this section we provide an argument that may weaken their conclusion. Before that, let us clarify in some detail what Einstein thought about the nature of the quantum state. It is well-known that he suggested that the $\psi$ function does not describe individual systems---as required instead by the definitions provided in the previous section---but \emph{ensemble of systems}. From several arguments raised against the completeness of QM, indeed, it emerges evidently that he favored a statistical interpretation of this theory. Interestingly, such arguments span almost his entire career, as for instance (but not only) in \cite{Einstein:1909, Einstein:1936, Einstein:1949, Einstein:1953}, showing that he endorsed the ensemble view consistently. To this regard, \cite{Bacciagaluppi2009} (p.\ 150) even claim that ``Einstein was arguably the founder of the statistical interpretation''.\footnote{Clearly Bacciagaluppi and Valentini do not refer to Born's statistical interpretation of $|\psi|^2$---which gives the probability amplitude for \emph{individual} measurement outcomes---but to the ensemble view. It is worth noting that for Born individual measurement results are in principle unpredictable with certainty due to the inherent indeterminism of the theory, and the indeterminate nature of quantum objects. On the contrary, Einstein interpreted this unpredictability as a sign of the incompleteness of QM. Indeed, as we will see in the remainder of this section, Einstein considered the statistical character of QM as the cause for its incompleteness. These two physicists, then, held opposite metaphysical views concerning the interpretation of quantum theory. The reader may refer to the Born-Einstein exchange of letters \citep{Born:1971}, since the issue of the interpretation of QM is discussed in several places.}

For spatial reasons, in the present essay we will take into consideration only three of them which we find particularly relevant: the first is contained in the reports of the fifth Solvay Congress held in 1927 which can be found in \cite{Bacciagaluppi2009}, the second appeared in Einstein's 1936 essay ``\emph{Physics and Reality}'', finally, the third is given in Einstein's replies to the essays written for his intellectual biography in 1949, contained in the well-known volume edited by Schilpp, \emph{Albert Einstein Philosopher-Scientist}.

At the fifth Solvay Congress Einstein proposed a simple argument aimed at showing the incompleteness of the quantum mechanical representation of individual systems. Let's consider the following physical situation: a beam of electrons is directed towards a screen $S$ where a small slit $O$ is present, behind the screen there is a hemispherical photographic film $P$ with a large radius. Some of these electron will pass though $O$, and will spread uniformly in all directions on $P$, given that the waves associated with their state of motion are diffracted at $O$. The probability that an electron hits the screen in a determinate point $x$ of the film is measured by the intensity at $x$ of the spherical waves after the diffraction at $O$. In virtue of the laws of QM, however, it will not in general be possible to predict wth certainty the exact locations where the particle will hit the film; indeed, only the probability that an electron will be found at a given point of space can be calculated. Einstein, then, compares two possible interpretations of quantum theory with respect to this example that substantially differ about the nature of the quantum state:
\begin{enumerate}
\item Interpretation I: the quantum state $\psi$ is not associate to a single electron, but to an ensemble of systems, a cloud of electrons. Consequently, the quantity $|\psi|^2$ represents the probability that at a given point $x$ of $P$ is found a particle of the statistical ensemble. Thus, quantum theory does not provide information about individual processes, ``but only about the ensemble of an \emph{infinity of elementary processes} (Einstein in \cite{Bacciagaluppi2009}, p.\ 441, emphasis added);
\item Interpretation II: The theory provides a complete description of individual processes, so that each electron directed towards $S$ can be fully accounted in terms of its quantum state---i.e.\ to every electron is associated a wave packet which is subject to diffraction at $O$: here $|\psi|^2$ represents the probability that the particle is found at the point $x$ of the film.
\end{enumerate}

According to Interpretation I, the quantum state represents an ensemble of electrons, and not individual systems. In this case the quantity $|\psi|^2$ represents the probability that in a particular point of the film there is \emph{a} particle of the cloud. Here we see a crucial feature of this statistical interpretation: it is silent with respect to individual processes. On the contrary, conforming to Interpretation II, the squared modulus of $\psi$ ``expresses the probability that at a given instant \emph{the same} particle is present at a given point (for example on the screen). Here, the theory refers to an individual process and claims to describe everything that is governed by laws'' (\emph{ibid}). In this second interpretation, $\psi$ provides a complete description of a single, specific electron. Einstein's argues in this manner against Interpretation II:
\begin{quote}
If $|\psi|^2$ were simply regarded as the probability that at a certain point a given particle is found at a given time, it could happen that \emph{the same} elementary process produces an action \emph{in two or several} places on the screen. But the interpretation, according to which $|\psi|^2$ expresses the probability that \emph{this} particle is found at a given point, assumes an entirely peculiar mechanism of action at the distance, which prevents the wave continuously distributed in space from producing an action in \emph{two} places on the screen (\emph{ibid}).
\end{quote}

\noindent In sum, Einstein claims that QM prevents multiple detections of a single electron on the screen in virtue of some mechanism of action at a distance---i.e.\ the collapse of $\psi$. However, he argues, such a mechanism is in contradiction with the principle of locality, and thus, with the theory of relativity. Since locality was a necessary physical principle according to Einstein's opinion, it follows that this action at a distance is avoided only claiming that the wave function does not provide a complete description of individual systems and individual processes. Notoriously Einstein favored Interpretation I, which speak in favor of an ensemble interpretation of the quantum state.

The second argument is taken from \cite{Einstein:1936}. In this essay Einstein explicitly claims that the incompleteness of quantum theory is a \emph{consequence} of the statistical nature of the quantum laws: ``the incompleteness of the representation'' of quantum systems he says, ``is the outcome of the statistical nature (incompleteness) of the laws [of quantum mechanics] (p.\ 374). Einstein, then, provides the following example to support his claim: consider a quantum state $\psi_r$ which corresponds to the periodic solution of the Schr\"odinger equation in order of increasing energy. Suppose that a system is in the lowest energy state $E_1$ represented by $\psi_1$, and that after a certain finite amount of time a small force acts on our system. Einstein notes that from the Schr\"odinger's evolution, one obtains the following wave function:$$\psi=\sum c_r\psi_r.$$ Then, he asks whether $\psi$ describes a real state of the system:

\begin{quote}
If the answer is yes, then we can hardly do otherwise than ascribe to this condition a definite energy $E$, and, in particular, such an energy as exceeds $E_1$ by a small amount (in any case $E_1< E < E_2$). Such an assumption is, however, at variance with the experiments on electron impact such as have been made by J. Franck and G. Hertz, if, in addition to this, one accepts Millikan's demonstration of the discrete nature of electricity. As a matter of fact, these experiments lead to the conclusion that energy values of a state lying between the quantum values do not exist. From this it follows that our function $\psi$ does not in any way describe a homogeneous condition of the body, but represents rather a statistical description in which the $c_r$ represent probabilities of the individual energy values [...] The $\psi$ function does not in any way describe a condition which could be that of a single system; it relates rather to many systems, to ``an ensemble of systems'' in the sense of statistical mechanics (\cite{Einstein:1936}, p.\ 375).
\end{quote}
\noindent In addition, at the end of his analysis, Einstein underlines that the quantum state provides such a statistical description for two essential reasons: one the one hand, the performance of a measurement introduces unknown and uncontrollable perturbation on the system, on the other, and more importantly, because the wave function does not represent individual systems, as shown in the above example. Thus, he concludes, the incompleteness of quantum mechanics is a consequence of the statistical nature of its laws.

Furthermore, in this essay Einstein briefly refers also to the EPR incompleteness argument, presenting it in form which is very close to that included in his letter to Schr\"odinger mentioned in the previous section. In this argument, a measurement of a two-valued operator performed on the system $A$---part of a complex entangled system $AB$---will affect the state of the system $B$ which is space-like separated. Then, it is possible to ascribe to $B$ multiple quantum states, which do not describe unambiguously the unique physical state of $B$. To this regard, Einstein notes it is sufficient to interpret $\psi$ as referring to an ensemble of systems in order to eliminate ``every difficulty''. Referring to this example Einstein claims: ``The fact that quantum mechanics affords, in such a simple manner, statements concerning (apparently) discontinuous transitions from one total condition to another without actually giving a representation of the specific process, this fact is connected with another, namely the fact that the theory, in reality, does not operate with the single system, but with a totality of systems.'' (\emph{ibid.}, p.\ 376). Interestingly, this statement clearly shows that Harrigan and Spekkens' argument aimed at showing that Einstein endorsed a $\psi-$epistemic view rests essentially on his statistical conception of the quantum state.

Finally, the third textual evidence for Einstein's endorsement of the statistical view is taken from his intellectual biography \citep{Einstein:1949}. In order to show the statistical nature of quantum theory, he considers the case of a radioactive atom whose average decay time is definite, and which is exactly localized in a point in space. It is well-known that conforming to the formalism of QM, the radioactive process of emission of a light particle is described by a three-dimensional wave function which is different from zero only in the small region occupied by the atom at the initial time $t_0$, but at successive instants $t$ it spreads in space. Such a function provides the probability to find the particle in given region of space if a measurement of position is performed, however, it does not contain any information concerning the exact time of the disintegration of the radioactive atom. Thus, Einstein asks whether the description of the decay process provided by quantum theory is complete. Not surprisingly, he answers in the negative:
\begin{quote}
The immediate plausible answer is: No. For one is, first of all, inclined to assume that the individual atom decays at a definite time; however, such a definite time-value is not implied in the description by the $\psi-$function. If, therefore, the individual atom has a definite disintegration time, then as regards the the individual atom its description by means of the $\psi-$function must be interpreted as an incomplete description. In this case the $\psi-$function is to be taken as the description, not of a singular system, but of an ideal ensemble of systems. In this case one is driven to the conviction that a complete description of a single system should, after all, be possible; but for such complete description there is no room in the conceptual world of statistical quantum theory (\cite{Einstein:1949}, p.\ 668).
\end{quote}
\noindent Einstein continues by saying that
\begin{quote}
Insofar, then, as a quantum-theoretician takes the position that the description by means of a $\psi-$function refers only to an ideal systematic totality but in no wise to the individual system, he may calmly assume a definite point of time for the transformation. But, if he represents the assumption that his description by way of the $\psi-$function is to be taken as the \emph{complete} description of the individual system, then he must reject the postulation of a specific decay-time (\cite{Einstein:1949}, p.\ 669-670).
\end{quote}
\noindent In conclusion, he claimed programmatically that if one aims at describing individual quantum systems with quantum theory, one is inevitably lead to ``implausible theoretical conceptions'' as we have underlined with the above examples. In particular, one is forced to accept a spooky  mechanism of action at a distance, which is in stark contrast with the principles of relativity theories. Thus, according to Einstein, the only meaningful reading of quantum theory is statistical, meaning that it describes only ensemble of systems, and not individuals (cf.\ \cite{Einstein:1949}, p.\ 671-672).\footnote{Referring to this, two facts should be mentioned: on the one hand, although for Einstein the statistical interpretation was the most meaningful reading of quantum theory, he was \emph{not} fully satisfied with the then current situation about the interpretation of QM. Thus, the ensemble view was just the ``least bad'' option available at the time, attending new developments of physics. To this specific regard the reader may refer to \cite{Einstein:1949}, pp.\ 671-673. On the other hand, Einstein was convinced that in order to provide a complete description of individual processes one should have added variables to the formalism of quantum theory. Alternatively stated, Einstein was in favor of a possible completion of quantum theory with the aid of hidden variables, or better, he thought that quantum theory should include in its description definite properties for quantum systems, as underlined also by \cite{Howard:1985}. In spite of this, he did not supported the hidden variable interpretation of QM provided by \cite{Bohm:1952aa}; indeed, in \cite{Einstein:1953} severe objects against Bohm's theory were presented (cf.\ \cite{Myrvold:2003} for a discussion). To this regard, the reader should also refer to Einstein's correspondence with Born \citep{Born:1971}, as for instance letter 99 dated 12th May 1952, where Einstein judged Bohm's theory as a ``cheap solution'' to the problems of quantum theory.}

After having showed Einstein's views on the statical nature of quantum theory---which is the source of its incompleteness---let us come back to Harrigan and Spekkens' argument. To state that Einstein endorsed a $\psi-$epistemic view of the quantum state, and simultaneously to acknowledge that he supported the statistical interpretation, imply that the latter must be considered a $\psi-$epistemic theory, dealing with observers' knowledge of the quantum state. 

Thus, a question immediately arises: How do Harrigan and Spekkens reinforce their conclusion that the ensemble view is a $\psi-$epistemic model? They answer it by saying that the notion of ``ensemble'' in Einstein's jargon is nothing but a way to talk about probabilities reflecting an observer's knowledge: 
\begin{quote}
the only difference between ``ensemble talk'' and ``epistemic talk'' is that in the former, probabilities are understood as relative frequencies in an ensemble of systems, while in the latter, they are understood as characterizations of the incomplete knowledge that an observer has of a single system when she knows the ensemble from which it was drawn. Ultimately, then, the only difference we can discern between the ensemble view and the epistemic view concerns how one speaks about probabilities, and although one can debate the merits of different conceptions of probability, we do not feel that the distinction is significant in this context, nor is there any indication of Einstein having thought so (\cite{Harrigan:2010}, p.\ 150). 
\end{quote}

Given the importance of this conclusion, let us discuss whether it is really sound. Let us see, then, if we can really affirm that the ensemble view is $\psi-$epistemic. In order to address these issues, it is useful to introduce the tenets of the statistical interpretation of QM relevant to our discussion. In what follows we will consider the statistical interpretation of QM contained in \cite{Ballentine:1970}, which is still today the most detailed presentation of this approach to quantum theory\footnote{For a more recent perspective on the statistical interpretation of quantum theory, the reader may refer to \cite{Bowman:2008}.} It is worth noting that Ballentine himself identified the statistical view with Einstein's interpretation of quantum theory; indeed, he explicitly writes that ``if we identify the Copenhagen Interpretation with the opinions of Bohr, then the Statistical Interpretation is rather like those of Einstein'' (\emph{ibid.}, p.\ 358). Let us then introduce the basic tenets of the statistical interpretation:

\begin{itemize}
\item Pure states provide a description of statistical properties of an \emph{ensemble} of similarly prepared systems. The quantum state $\psi$ does not need to completely describes individual systems;
\item Quantum states formally represent the set of certain state preparation procedures. Thus, systems which undergo the same preparation procedures for certain observable will have similar properties, but they are not equal in all their respects---they may differ with respect to those properties for which they are not subjected to the same preparation procedures (The uncertainty principle means that is not possible to obtain an ensemble of identically prepared systems \emph{for all their properties});
\item A certain eigenstate of a given observable represents the ensemble of those systems which are in that particular state;
\item Generally, quantum theory does not predict the result of individual measurement of a certain observable. However, the probability of each possible outcome can be verified by the iterated repetition of state preparations and measurements, eventually constructing the statistical distributions of the results;
\item Generally, quantum theory predictions are not relevant for individual, single measurements. Quantum mechanical calculations rather pertains to an ensemble of similar measurements;
\item According to the ensemble view, probability refers to ``the \emph{relative frequency} (or measure) of the various eigenvalues of the observable in the conceptual infinite ensemble of all possible outcomes in the identical experiments (the sample space)'', whereas the \emph{statistical frequency} refers to the ``results in an actual sequence of experiments. The probabilities are properties of the state preparation method and are logically independent of the subsequent measurement'' (\cite{Ballentine:1970}, p.\ 361, italics added);
\item In the statistical view, probability does not refer to the observer's knowledge. Referring to this, Ballentine is explicit: ``[i]n contrast to the Statistical Interpretation, some mathematicians and physicists regard probability as a measure of knowledge, and assert that the use of probability is necessitated only by the incompleteness of one's knowledge. This interpretation can legitimately be applied to games like bridge or poker [...]. But physics is not such a game, and as Popper has emphasized, one cannot logically deduce new and verifiable knowledge---statistical knowledge---literally from a lack of knowledge'' (\emph{ibid.}).
\end{itemize}

It is easy to recognize these tenets in Einstein's view of QM. In the first place, states also here refer to ensembles, thus, they do not describe individual systems. In the second place, the predictions of quantum theory do not concern individual processes, but only statistical distributions of measurement results. Finally, the quantum probabilities do not refer to observers' knowledge of particular experimental situations. Hence, it is fair to establish a strong theoretical continuity between Ballentine's presentation of the ensemble view and Einstein's interpretation of quantum mechanics. This fact, as we will see, will be crucial in our argumentation.

In the first place, there is a (ontological) crucial difference between the ensemble view and the requirements imposed to ontological models by Harrigan and Spekkens. According to the statistical interpretation, the quantum state does not \emph{aim} to represent individual systems, meaning that it does not tell us anything about single, individual quantum systems. Neither does it tell us anything about individual measurements. Everything that can be said about quantum systems concern sets of similarly prepared objects. It is worth noting, however, that this view does not fail to describe individual systems, since these just lie outside its scope. Unfortunately, this crucial aspect of the ensemble view is not taken into account by Harrigan and Spekkens: to claim that it is a $\psi-$epistemic model because $\psi$ does not describe individual states, it is to assume that this view somehow failed to describe individual systems. However, this is simply to ignore the relevance of ensembles for the statistical interpretation. Referring to this, it is interesting to note that Ballentine carefully distinguishes between two different types of interpretations of QM: the statistical view (according to which the quantum state does not provide a description of individual systems---as repeatedly stressed by Einstein) and interpretations in which pure states provide descriptions of individual systems. Harrigan and Spekkens never make this distinction clear, and they seem to be assessing the statistical view in terms of its success at describing individual systems, which, as we showed, is not in the business of such a view. Alternatively stated, Harrigan and Spekkens evaluate the statistical interpretation with the same criteria used to evaluate completely different sort of models, i.e.\ those in which the quantum state refers to individual systems, putting the ensemble view in the wrong category.

In the second place, another crucial point to highlight is that the ontic space of the statistical interpretation is not one of individuals, but of ensembles. This allows for an alternative reading of the ontic state: it provides a complete description of the properties of an ensemble, not of individuals. And there is nothing else to know about ensembles that is not provided by the quantum state. The upshot of the present discussion is that the sort of $\lambda$ that the statistical interpretation poses is completely different in nature with respect to that employed by Harrigan and Spekkens. Moreover, also the relation between $\lambda$ and $\psi$ should be assessed differently within the statistical view. These facts are completely overlooked in Harrigan and Spekkens' account. This justifies one of our claims against their classification: it is too narrow and puts severe constraints on the structure of the ontic spaces, to the effect that even an interpretation where observers' knowledge does not play any role is classified as epistemic. The crucial issue here is that the assumption according to which the ontic state exclusively represents the complete state of an individual system does not find any theoretical justification within the ensemble view, since it is fundamentally not about individuals, but about sets of identically prepared systems which are completely described by $\psi$.

Naturally, the way to amend Harrigan and Spekkens' misrepresentation of the statistical interpretation is to consider the possibility to enlarge the description of ontic states and ontic spaces to ensembles. This amounts to allowing the ontic space to describe either individual systems or ensemble of systems. Then, $\psi$ may be assessed in terms of its success in capturing either ontic states' features. When it comes to the statistical view, $\psi$ must be assessed in relation to an ontic state of ensembles. It happens that so interpreted it says all can be said about such an ontic state.\footnote{Here we are not claiming that the statistical view provides an ultimate description of physical objects, resolving once and for all the ontological disputes of QM. Indeed, both Einstein and Ballentine conceive the possibility to complete the description of systems provided by this interpretation. However, both these physicists say that as far as \emph{quantum mechanics} is concerned, the statistical view is the most detailed approach at our disposal if one wants to avoid theoretical and interpretational difficulties.} In accordance with this, the ensemble view would be a $\psi-$complete model, which following Harrigan and Spekkens' definition, cannot be $\psi-$epistemic. Therefore, the assumption that the authors made defining what ontological models are simply leaves out the possibility to evaluate correctly the status of the quantum state in the context of the statistical interpretation.

There is a further argument that concerns probabilities. As we have seen from the principles of the ensemble view, it is not true that probabilities in such an interpretation represent observers' knowledge, as Harrigan and Spekkens maintained. Indeed, in their essay there is no theoretical explanation for which probabilities in the ensemble view should be understood as observers' incomplete knowledge. While it is true that in the ensemble view quantum  probabilities refer to relative frequencies, as the authors claim, they nonetheless refer to ``the relative frequency (or measure) of the various eigenvalues of the observable in the conceptual \emph{infinite ensemble of all possible outcomes in the identical experiments} (the sample space)'' \cite{Ballentine:1970}, p.\ 361, emphasis added). It is the statistical frequency that refers to actual sequence of measurement performed on a particular ensemble of quantum systems. Thus, the authors are conflating relative frequencies, that is, quantum probabilities, with mere statistical frequency. This fact comes from a misrepresentation of the notion of ensemble as used in the statistical view. Indeed, the authors claim that the ``ensemble talk'' refers solely to the fact that probabilities should be interpreted as relative frequency in an actual, particular ensemble of systems. However, Ballentine reminds us that this is not the case: the notion of ``ensemble'' at play in the statistical view is quite different and deeper, since it refers to the infinite conceptual ensemble of similarly prepared system; ``for example, the system may be a single electron. Then the ensemble will be the conceptual (infinite) set of all single electrons which have been subjected to some state preparation technique (to be specified for each state), generally by interaction with a suitable apparatus'' (\emph{ibid.}). This is also underlined by Einstein himself, when he claimed that the quantum state refers to ``an infinity of elementary processes'' and to ``an ideal ensemble of systems'', as seen a few lines above. In addition, Ballentine clearly states that probabilities in the statistical interpretation are inherently associated with states preparation procedures, and are metaphysically and logically independent from actual measurements---in this latter case we deal with statistical frequencies. Thus, it is false to claim that, in the context of the ensemble view, probabilities refer to relative frequencies concerning actual measurements performed on particular sets of systems. If this is true, then the supposed equivalence between ``ensemble talk'' and ``epistemic talk'' has no theoretical basis. As a consequence, it would be misleading to consider the ensemble view as an $\psi-$epistemic model, since observers' knowledge plays absolutely no role in this context.

In conclusion, in this section we argued that there is no theoretical basis to consider the statistical interpretation of quantum theory a $\psi-$epistemic model, \emph{contra} Harrigan and Spekkens' claims. The reason lies in a difference between ontic states in the ensemble view and in Harrigan and Spekkens' classification: they represent ensembles in the former, but individuals in the latter. Moreover, there is an issue with respect to the interpretation of probabilities: the authors do not provide sound arguments supporting the thesis for which probabilities in the ensemble view should refer to incomplete observers' knowledge. Consequently, claiming that the statistical view entails an epistemic interpretation of the quantum state would require further argumentation. This naturally puts into question the conclusion that Einstein defended a $\psi-$epistemic view as well. Of course, the authors may argue that Einstein did not endorsed what is now called the ensemble view, but a different sort of epistemic interpretation. However, this should be proved with more rigorous argumentation, especially in the light of the textual evidence provided in this section according to which Einstein did endorse the ensemble view, and that his thoughts are faithfully represented by \cite{Ballentine:1970}.

\section{On the Perspective Independency of $\lambda$}
\label{4}

In the previous section we argued that the assumption that $\lambda$ applies to \emph{individual} systems cannot capture the basic tenets of the statistical interpretation of quantum mechanics. In this section, we will analyze and discuss a second assumption in Harrigan and Spekkens' classification with respect to the nature of $\lambda$, namely, that it is taken to be independent of any reference system.\ We will call this assumption \emph{perspective independence}, according to which  quantum systems instantiate absolute, intrinsic properties. Though intuitive at first sight, we will argue that the perspective independence assumption is at odds with some interpretations of QM that adopt a \emph{relational} (or \emph{perspectival}) ontology---\emph{the Modal-Perspectival Quantum Mechanics} (PQM henceforth) and \emph{Relational Quantum Mechanics} (RQM henceforth). This implies that some ontological models of QM would not fit into Harrigan and Spekkens' classification, since they diverge about the nature of $\lambda$.

\subsection{$\lambda$-Perspective Independent?}
\label{4.1}

Let us begin by introducing in detail the perspective independence assumption. As recalled in the previous sections, Harrigan and Spekkens begin their characterization of $\lambda$ by claiming that the primitive notions of ontological models are properties of microscopic systems: 

\begin{quote}
[a] preparation procedure is assumed to prepare a system with certain properties and a measurement procedure is assumed to reveal something about those properties. A complete specification of the properties of a system is referred to as the ontic state of that system, and is denoted by $\lambda$ (\cite{Harrigan:2010}, p.\ 128).
\end{quote}

As we mentioned previously, one of the criteria to sorting ontological models out consists in specifying the type of relation existing between $\lambda$ and $\psi$. Note that \emph{$\psi$-incomplete} models take $\psi$ to be just a partial description of $\lambda$, in the sense that the latter involves more properties than those captured by $\psi$. This \emph{failure} (\emph{ibid.}, p.\ 131) may require $\Lambda$ either to be parameterized by $\psi$ \emph{and} by supplementary variables, or to be not parameterized by $\psi$ whatsoever, which renders $\psi$ purely epistemic.

Be that as it may, we claim that Harrigan and Spekkens' characterization of quantum ontological models requires $\lambda$ to meet the following assumptions:
\begin{itemize}
\item \emph{Intrinsicality}: a quantum system instantiates \emph{intrinsic properties} that may be revealed through measurements.
\item \emph{Absolute objectivity and completeness}: $\lambda$, the objective, complete description of a quantum system, is primarily given by univocally specifying the intrinsic properties of a quantum system---such properties hold absolutely, i.e.\ for any possible perspective.
\end{itemize}
In virtue of these two assumptions, it can be said that in $\psi$-ontic models $\lambda$ `encodes' the $\psi$, for $\lambda$ is only consistent with one choice of $\psi$. If a quantum system only instantiates intrinsic properties, then any description of it will consist in providing \emph{monadic predicates}, that is, one-place logical predicates like $F(x)$. 

Consequently, a complete description of a quantum system will consist in providing an exhaustive list of monadic predicates that may be ascribed to it univocally for any perspective. It is important to highlight \emph{univocally} here: one of the most salient features of intrinsic properties is that things possess them in virtue of the way themselves are, regardless of their relations with external things (\cite{Lewis:1986}, p.\ 61). Thus, it follows that if $\lambda$ is a complete description of the intrinsic properties of a quantum system, any consistent $\psi$-complete model is forced to provide only one $\psi$.  
To explain this point more clearly, we can think of the opposite scenario, when we have two different $\psi$s for the same $\lambda$. These $\psi$s might be regarded as two different perspectives, representing \emph{overlapping} probabilities, in the sense that there exists a pair of preparation procedures, $P_{\psi}$ and $P_{\phi}$, such that $p(\lambda | P_{\psi})p(\lambda|P_{\phi})\neq0$. Hence, $\lambda$ does not encode $\psi$, which translates into a lacking of knowledge about $\lambda$. On the contrary, if $\lambda$ is only consistent with one choice of $\psi$, it follows straightforwardly that it is consistent with one of the perspectives from which a quantum system is described.\ In consequence, as there is only one choice of $\psi$ compatible with $\lambda$, it is independent of the perspective we choose---we will trivially choose the right one.\ This aspect reveals a third implicit assumption that in turn relies upon the previous ones:
\begin{itemize}
\item \emph{Perspective independence}: A complete description of the intrinsic properties of a quantum system can be only given by one choice of $\psi$.
\end{itemize}

Alternatively stated, there is \emph{only one} correct perspective to describe the state of a quantum system. Hence, intrinsicality, absolute objectivity and completeness and perspective independence naturally leads to categorize overlapping probability distributions as \emph{epistemic} states: such probability distributions must be read as \emph{ignorance} about the intrinsic properties a quantum system instantiates. In other words, if two $\psi$s can be ascribed to some $\lambda$, then they are equally failing to capture a quantum system's intrinsic properties.

These assumptions are common in the literature on quantum physics and have roots in classical physics. If a classical coin shows `head' after flipping it, then we assume the coin really instantiates the property of ``being head'' \emph{regardless} the perspective from which the property ``being head'' was registered. If we were asked whether the coin landed `tail' after flipping it, we would somehow know that an objective, complete answer will be given by a yes-no proposition. Any response lying in between must be considered as incomplete, in the sense of lacking the required knowledge to provide a yes-no answer. \emph{Mutatis mutandis}, these assumptions play an akin role when it comes to quantum physics within Harrigan and Spekkens' classification: $\lambda$ would ideally give us an objective, complete description of a quantum system by listing its (intrinsic) properties. 

This conventional wisdom can be found in different places in the standard literature, but for spatial reasons here we consider just two noteworthy sources.\ On the one hand, in their book \emph{The Quantum Theory of Measurement}, Paul Busch, Pekka Lahti and Peter Mittelstaedt introduce an \emph{objectification requirement} in order to know how definite properties are to be represented in the quantum formalism. Conforming to such a criterion, a property of a system is definite and objective if and only if the system's quantum state is a mixture, in the ignorance sense, of eigenstates of the observable corresponding to the property in question (\cite{Busch:1996}, p.\ 21). The requirement basically says that a quantum system instantiates an objective property if there is a probability equal to 1 of finding the system in an eigenstate of the observable representing such a property upon measurement. As Dennis Dieks notes, in such a mixture ``the presence of different pure states as components reflects our lack of knowledge about which one of these states actually obtains'' (\cite{Dieks:2009}, p.\ 761). The objectification requirement thus presupposes that quantum systems must really instantiate some properties and lack others \emph{univocally}---any failure of completely capturing such properties must be interpreted as ignorance. 
 
On the other hand, in their 2012 celebrated paper, Matthew Pusey, Jonathan Barrett and Terry Rudolph poses a no-go theorem for any model in which the quantum state ($\psi$) just represents mere information about an underlying physical state. Though the term `physical state' is not perfectly accurate, they claim that quantum systems have real physical states that are objective and independent of any observer. According to this view, physical properties are constant functions over the physical state and a quantum system genuinely instantiates a physical property \emph{only if} for every pair of distributions, they are disjoint. To put it differently, the existence of overlapping probability distribution would denounce a failure in the assignment of physical properties to quantum systems (\cite{Pusey:2012}, p.\ 1-2), which is in perfect agreement with Harrigan and Spekkens' view.

Any challenge to this attitude requires a radical revision of our understanding of properties and how we know them. In the longstanding debate about how the quantum formalism should be interpreted, some have proposed such a radical revision in the last decades. Two clear instances of this are PQM, mainly defended by Dennis Dieks (cf.\ \cite{Bene:2002}, \cite{Dieks:2009}, \cite{Dieks:2019}), and RQM, actively promoted by Carlo Rovelli (cf.\ \cite{Rovelli:1996}). Although different, both views rely on a strong relational ontology for quantum physics, according to which the properties of quantum systems are not absolute or intrinsic. In accordance with this, any description of physical systems must be given in terms of \emph{relational properties}, expressed through at least \emph{dyadic} predicates involving not only the target system \emph{S}, but also the reference system \emph{R} from which it is said that \emph{S} instantiates the property $F$. Thus, any description will look like a two-place predicate, $F(S, R)$, which reads ``$S$ is $F$ from $R$'s perspective''. In the following, we will briefly introduce both interpretations, focusing on their ontological tenets.

\subsection{Towards a Perspectival and Relational Quantum Mechanics}
\label{4.2}

PQM belongs to the class of the so-called modal interpretations of quantum mechanics (cf.\ \cite{vanFraassen:1991}, \cite{Dieks:1998}, and \cite{Lombardi:2017}). From a broad viewpoint, the modal interpretations are a significantly large family of interpretations of the quantum formalism holding at least three theses:

\begin{itemize}
\item \textbf{Non-collapse dynamics}: The quantum state evolves always unitarily according to some Schr\"odinger-type equation of motion;
\item \textbf{A single world}: there only exists a single actual world;
\item \textbf{Modal possibilism}: The quantum state (specifically, the dynamical state) represents what may be the case, that is, it represents \emph{real possibilities}.
\end{itemize}

Taking as an example a one-particle system in a superposition of \emph{z-}spin, $|\psi\rangle=\sqrt{\frac{1}{2}}(|\uparrow\rangle+|\downarrow\rangle)$, each term in the superposition represents different possibilities that might be actualized through some actualization rule. Such a superposed state never collapses, but evolves unitarily according to some Schr\"odinger-type dynamical equation. And here it comes the philosophical novelty of modal interpretations: the quantum reality is not exhausted by what is actual (that is, by which term we will eventually obtain in the actual world), but it also involves what is possible. Although supporters of the modal interpretation hold that there is only one actual world, wherein some set of self-adjoint Hermitian operators will acquire definite values, they also hold that there is a realm of unrealized possibilities.\ This proposal, however, should be clearly distinguished from the Many-World interpretation, since the latter holds that any term in a superposition corresponds to some actual existing world (or ``branch'' of the universal wave function), whereas the former holds that only one term corresponds to some actual state in our world.

One of the central debates within modal interpretations is about whether there is an \emph{a priori} privileged set of commuting observables. Conforming to some modal versions, a realist non-collapse interpretation is committed to selecting a privileged set of definite-valued observables out of all observables. Which is such a set has been one of the central issues of modal interpretations, raising several disputes about the nature of the actualization rule and observables.\ For instance, a Bohm-like interpretation may be seen as a modal interpretation in which the privileged observable is \emph{position}, so that the set of definite-valued observable for any quantum system will be given by all the observables that commute with position (cf.\ \cite{Bub:1997}). Others have picked out the \emph{Hamiltonian} as the privileged observable (cf.\ \cite{Lombardi:2008}). Since the Hamiltonian commutes with all the Casimir operators of the Galilean group (when the system is described in the reference frame of its center of mass), it plays a paramount role in building the quantum theory (cf.\ \cite{Lombardi:2010}). 

However, other modal interpretations disagree on selecting a privileged set of commuting observables \emph{a priori}. Instead, they hold that which subset of observables will acquire actual definite values depends on the form of the quantum state. Importantly, and this is the novelty with respect to more orthodox approaches, insofar as the form of the quantum state changes with time, such a subset of observables will also change with time. 

Putting aside these differences, we can claim that for these approaches \emph{the quantum state will depend upon the perspective from which the quantum system is given.} This idea has been championed by PQM, a \emph{perspectivalist} view of quantum mechanics within modal interpretations, supported mainly by Dennis Dieks and Gyula Bene.\footnote{For a discussion of relativistic quantum states in PQM the reader may refer to \cite{Myrvold:2002}.} According to this approach, the nature of quantum reality is inherent relational in so much as the properties of quantum systems are not intrinsic.

This chiefly means that any description of the state of a quantum system can be only given by relational properties, defined with respect to another physical system serving as `witness' (borrowing Kochen's expression, \cite{Kochen:1985}) or `reference system' (\cite{Bene:1997}). In this context, a quantum system $S$ can thus be represented by different quantum states relative to different reference systems. This fact entails that the same quantum system $S$ may have definite values of certain observables from the perspective of a reference system $R$, \emph{and} may be in a superposition with respect to those observables from a distinct reference system $O$. The upshot of all this is that as the nature of quantum states is \emph{intrinsically relational}, there is no matter of fact out of which we can privilege one perspective over the other. In this sense the nature of the quantum state is said to be \emph{perspectival}. 

In metaphysics, relational properties are those whose instantiation by some individual depends on the sort of relations that such an individual holds with other systems or its surrounding. What is important to remark about relational properties is what resources we need to exhaustively specify them. As already mentioned above, intrinsic properties are given by one-place predicates as, for instance, the property of having a certain mass, which is expressed by the proposition $p$, ``the body $c$ has a mass $=x$'', $p:M(c)=x$. In this case, we only need one placeholder to define the property and to ascribe truth values to the proposition $p$, $V(p)=1(0).$ Extrinsic properties behave quite differently. To begin with, they are given by at least two-place predicates, that is, two placeholders are required to define the property adequately. For instance, the property of ``having a weight $= x$'' remains undefined until the \emph{location} of an individual not be specified. In other words, the property is relational because requires to provide the whereabouts of an individual for the proposition in which it intervenes to be given with truth values. The weight of an apple will vary depending on whether it is on the Earth or on Mars, so one of the placeholders in the predicate function will be filled by an individual's location, (\emph{l}), $(M(\emph{o},\emph{l})=x)$, where \emph{o} is the apple and \emph{x} is weight of the apple according to the location \emph{l}. Note that for a relational property as ``having a certain weight'' the individual's location is essential in order to define meaningfully the property itself and cannot be eliminated.

In general, we intuitively think of physical systems (and things broadly conceived) as having both intrinsic and relational properties. Sometimes we want to talk about how a physical system is in itself, and thereby we rely on its intrinsic properties. However, sometimes we rather want to talk about a physical system in relation to something external to it, relying on its relational properties. Notwithstanding this, PQM claims that when it comes to quantum physics, we have both principled and empirical reasons to believe that quantum reality only involves relational properties. This primarily means that in order to specify the state of a quantum object, we must provide information concerning the reference system from which we are specifying such a quantum state (a \emph{perspective} in the jargon of PQM). Then, the quantum state is relational in exactly the same way as the property ``having a weight $= x$'' is defined in relation to a location. Let us see a concrete example.

PQM's starting point is generally the standard Hilbert space formalism with only unitary time evolution. In this framework, the largest quantum system is the universe as a whole, whose quantum state is a pure vector state $|\psi\rangle$. However, the truly relational nature of quantum systems comes up when we focus on subsystems of the whole universe. In these cases, quantum states will be represented by density operators acting on the Hilbert space of the subsystem $S$. In the standard representations, we give the quantum state of a system when we simply specify its density operator, $\rho$. But in PQM the quantum state is relational, meaning that its specification must be given by a two-place predicate. This can be represented in the formalism by attaching indices to the density operator, $\rho^S_R$, which reads ``the quantum state of subsystem $S$ relatively to $R$'', or ``from the perspective of $R$''.

There are two noteworthy cases. The first one is when $S$ coincides with $R$. The situation can be rephrased in terms of ``the quantum state of $S$ relatively to itself''. In this case, the quantum state of $S$ is given by partial tracing the quantum state of the universe (which by definition is in the pure quantum state $\rho^U_U= |\psi\rangle\langle\psi|$). 
Thus, the density operator for a subsystem $S$ relative to itself is $\rho^S_U=Tr_{(U\backslash S)}\rho^U_U=Tr_{(U\backslash S)}|\psi\rangle\langle\psi|$. If there is no degeneracy, then $\rho^S_S= |\psi\rangle\langle\psi|$.
In accordance with the modal tradition, the state $\rho^S_S$ codifies the physical properties that $S$ actually has. In other words, all the properties that are derived from the quantum state of $S$ from its own perspective are properties that $S$ possess on its own. However, the perspectival nature of quantum mechanics arises when the quantum state of $S$ is considered from an arbitrary external reference system, $\rho^S_R$. A particular case of this is when the system $S$ is contained in a bigger system $A$. To obtain the relational state of $S$ from $A$ we trace out the degrees of freedom of $A$ that do not belong to $S$. Then, $\rho^S_A$ can be defined as the density operator $\rho^S_A=Tr_{(A\backslash S)}\rho^A_A$.

Let us stop at this point to make some comments. The relational nature of the quantum state means that the state and physical properties of any quantum system crucially (and we would say \emph{by definition}) depend upon specifying a reference system with respect to which they are given. Importantly, this means that the relational nature of the quantum state is \emph{irreducible}, in the sense that the state cannot be boiled down to a non-relational specification of the quantum state. This suggests, in turn, that the relational nature of the quantum state does not represent an epistemic limitation in our way of ascribing states to quantum physical systems, as if there were some properties we failed to capture. Contrarily, this reflects an intrinsically relational quantum reality. When we say that a quantum system is in a superposed state of spin from a reference system $R$, but in a definite state, say $|\uparrow\rangle$, from a reference system $A$, we are not saying that $R$ does not really know that $S$ is in a  $|\uparrow\rangle$ state, but that from $R$'s perspective $S$ is indeterminate in spin. Such a relational state is as real as any other relational state. 

This might strike someone as perplexing and paradoxical: We have two opposing descriptions of the same situation that turn out to be on equal footing. However, according to PQM, this ``paradox'' only follows from not taking seriously the quantum mechanical formalism and its relational nature.\footnote{Nonetheless, this situation is pervading in physics and many examples can be given. For instance one may think about velocity in classical physics: such a quantity is always defined in relation to a particular reference system, however, this does not turn velocity into a subjective magnitude. It is just \emph{relative} to such a reference frame.} If quantum systems can only instantiate relational properties, any fair description ought to be given by relational states. In other words, in the quantum ontology there are not intrinsic properties whatsoever, thus, any description resorting on monadic predicates, or reducing relational properties to intrinsic ones, is just off the right track. This is also valid for those cases where the system and the reference system are the same. It could be argued that quantum systems actually instantiate intrinsic properties, namely, those defined from the perspective of the system itself. Putting aside quantum mechanical reasons to reject this argument (cf.\ \cite{Dieks:2019}), the argument is conceptually misguided: the state of a system with respect to itself is just one perspective among many others, being merely a special case of a relational property. The fact that those properties be said to be intrinsic is merely linguistic, not grounding any metaphysical claim.

Let us now turn to a more radical relational ontology for quantum mechanics. Rovelli's RQM shares to a good extent PQM's tenets, though it takes even a step further in the relativization of the quantum state.\ RQM begins by claiming that quantum mechanics is not actually about the quantum state ($|\psi\rangle$), as at first glance it seems, but about values of physical variables. The quantum state is just ``a fictitious nonphysical mental construction'' (\cite{Rovelli:1996}, p.\ 1645), which serves as a bookkeeping device for storing the information about the history of a system's interactions. This is one of the most notable differences between PQM and RQM: whereas in the former view the relational quantum state truly represents the relational nature of the quantum world, in the latter the quantum state is irreducibly epistemic. What is in fact objective in the relational quantum world envisaged by Rovelli is the real events or processes brought about in the course of the interactions between physical systems. In RQM,
\begin{quote}
[t]he real events of the world are the ``realization'' (the `coming to reality', the `actualization') of the values of $q, q'', q''', \dots$ in the course of the interaction between physical systems. This actualization of a variable $q$ in the course of an interaction can be denoted as the quantum event $q''$ (\cite{Dorato:2016}, p.\ 243).
\end{quote}

\noindent Hence, the micro-world described by the quantum mechanical formalism is made up of processes that characterize sequences of quantum events, that is, physical systems whose properties take values due to interactions. However, the values that properties take for a system are not absolute and intrinsic, but \emph{relational}. As in PQM, that a system's property $Q$ acquires the value $\emph{q}$ does not mean that $\emph{q}$ is an absolute value, but it is relative to an external reference system $R$ with which $S$ has interacted.  What the quantum mechanical formalism shows is that different physical interactions (for instance, agents performing measurements in different laboratories, or particles colliding spontaneously in the surroundings of a faraway star) can deliver different accounts of a quantum event. For the same token, it cannot be said that an isolated system has definite value for some variable $Q$: only interactions allow a system to take definite values. Insofar as all interactions stand on equal footing, every interacting physical system can be taken as defining a perspective from which some system's property acquires a definite value, which goes along with PQM's spirit.\footnote{The metaphysical implications of RQM has not been yet in-depth investigated. Two exceptions are \cite{Dorato:2016} and \cite{Candiotto:2017}.} 

\subsection{$\lambda$ in a Relational Quantum Land}
\label{4.3}

Both PQM and RQM radically change our conventional understanding of how physical systems should be identified through their properties.\ What consequences do these views entail for Harrigan and Spekkens' classification? In the first place, it is clear that the authors' assumptions about the nature of $\lambda$ are completely at odds with those of PQM and RQM. In consequence, the overarching relational attitude of both views imposes criteria to characterize $\lambda$ that diverge greatly with respect to those of Harrigan and Spekkens’. In the second place, PQM and RQM in a derivative sense allow two quantum states to correspond to overlapping probability distributions without being committed to consider $\psi$ to be epistemic. In the third place, both views bring up a different notion of objectivity, which clashes with that assumed by Harrigan and Spekkens. 

In subsection 4.1, we mentioned that their classification takes for granted three assumptions with respect to the nature of the ontic state, namely, \emph{intrinsicality, absolute objectivity and completeness}, and \emph{perspective independence}. Naturally, from the point of view of RQM and PQM, these assumptions can no longer be accepted. Within these interpretations, quantum systems do not possess intrinsic properties at all, but rather instantiate relational ones. Thus, it follows that $\psi$, if it is meant to represent (partially or completely) the quantum reality, must be relational due to principled metaphysical reasons. Hence, if intrinsicality is an assumption about what kind of properties a complete description of a quantum system must involve, PQM and RQM categorically reject it, holding a different principle:

\begin{itemize}
\item \textbf{Relationality}: a quantum system instantiates \emph{relational properties}, which require a reference system to be meaningful.
\end{itemize}

Under this assumption, $\psi$ no longer relates to a $\lambda$  involving intrinsic properties, but to a $\lambda$ involving properties always \emph{relatively to} a reference system. 
Clearly, the nature of $\lambda$ will also be relational. For instance, in the context of RQM, since a quantum event is defined by the interactions between a physical system $S$ and a reference system $R$,  $S$ acquires certain values uniquely with respect to $R$. A third system $O$, with which $S$ and $R$ do not interact, can rightfully offer a description of $S$ completely different with respect to that of $R$. 

Taking seriously into account the relational nature of quantum systems, $R$ and $O$ refer to the ``same'' quantum event---the interaction between $S$ and $R$. However, this quantum event may receive at least two different descriptions insofar as these refer to two different reference systems---one of them has interacted with $S$, whereas the other does not. The point is that there is nothing missing here for either perspective: both descriptions are on equal footing and just express the relational nature of quantum reality.

By rejecting intrinsicality, RQM and PQM deny the assumption of absolute objectivity and completeness as well. In the case of RQM, when two different reference systems $R$ and $O$ interact differently with a physical system $S$, they represent two equally valid perspectives from which the system can be characterized. Along the same line, from PQM's view, as there only exist relational properties, each perspective defines its own objective and complete set of (relational) properties of the system in relation to itself. Hence, the notion of completeness of the description can no longer be absolute, but relative:

\begin{itemize}
\item \textbf{Relative objectivity and completeness}: an objective, complete description of a quantum system's properties is given by the relational properties that a quantum system instantiates in relation to a reference system.
\end{itemize}

An important result following from these considerations is that quantum mechanics, according to RQM and PQM, makes it plausible to objectively ascribe \emph{more than} one state to the same physical system $S$. In other words, two distinct reference systems $R$ and $O$ can ascribe different $\psi$s to the same $S$, because such $R$ and $O$ will specify different complete sets of relational properties for the same quantum system. This directly refutes the identification between providing overlapping probability distributions and \emph{ignoring} which is the set of complete properties of a quantum system. In other words, RQM and PQM make automatically false any inference aiming at showing that providing two different $\psi$s for a given $\lambda$ leads to a purely epistemic reading of such $\psi$s. This is particularly interesting because it paves the way to regard overlapping probability distributions as \emph{$\psi$-ontic}, blurring out the distinctive feature of \emph{$\psi$-epistemic} ontological models. One of the lessons we can take from this is that being \emph{$\psi$-epistemic} not only requires that there is possible to provide overlapping probability distributions for a given $\lambda$, but also that $\lambda$ satisfies \emph{intrinsicality, absolute objectivity and completeness} and \emph{perspective independency}.

It is worth noting a subtle difference between PQM and RQM at this point. Whereas PQM would take a relational $\psi$ to be \emph{$\psi$-complete}, and thereby \emph{$\psi$-ontic}, RQM's attitude is manifestly instrumentalist with respect to $\psi$. For PQM, it is clear that any fair description of an irreducibly relational world must be given by relational quantum states, which by definition include the possibility of ascribing different states to the same physical system. This fact just reflects the relational nature of the world, and cannot be regarded as hampering a realist attitude. For RQM, by contrast, the quantum state is merely a useful tool for calculation and prediction, and because of this it is \emph{$\psi$-epistemic}. However, its epistemic or instrumentalist nature is not due to the possibility of having overlapping probability distributions, as it would follow from Harrigan and Spekkens' classification, but it is due to more principled reasons. RQM holds a weak realist attitude with respect to its sparse event ontology, which is of course also relational, but wherein $\psi$ is left out of the picture. Thus, the possibility of having different descriptions for a physical system (that is, of ascribing different values to physical variables) does not necessarily imply that there are some properties that our descriptions are not capturing. Neither does it imply that such descriptions are subjective: it is clear that the acquisition of definite values is a mind-independent physical fact, largely due to the existence of interactions among systems.

All these reasons naturally lead to finally rejecting \emph{perspective independence}

\begin{itemize}
\item \textbf{Perspective dependence}: A complete description of the properties of a quantum system can only be given by a relational $\psi$, which is specified in relation to a reference system.
\end{itemize}

To sum up, one of Harrigan and Spekkens' assumptions is that $\lambda$ is a complete description of reality independently of any reference system. This assumption was shown to be at odds with some current interpretations of QM that take $\lambda$ to be relational and perspectival. A relational metaphysics for QM does not only change the nature of quantum systems in relation to the properties that they may instantiate, but also the relation that $\lambda$ holds with $\psi$. Any adequate description of a perspectival quantum reality is forced to involve two-place predicates that refer, too, to the perspective from which a quantum system is described. Therefore, the assumption of a free-perspective $\lambda$ simply postulates a divergent metaphysics with respect to that of RQM's and PQM's, which reveals that the classification provided by these authors is too narrow to include them.

\section{Conclusion}
\label{5}

In this essay, we have argued that Harrigan and Spekkens' classification cannot be employed in order to convincingly accommodate some current interpretations of quantum mechanics. In particular, we have shown that interpretations promoting an ensemble view or a perspectival and relational ontology for the quantum realm severely conflict with some assumptions of Harrigan and Spekkens'  scheme. We provided a two-fold argumentation. In the first place, it has been shown that Harrigan and Spekkens' classification assumes that $\lambda$ is the complete description of single, individual quantum systems that are also perspective independent. On an operational basis, both assumptions set up an ontology according to which the basic constituents of \emph{every} quantum theory are individuals instantiating intrinsic properties. Conforming to it, $\psi$ was assessed in terms of its relationship and adequacy with such an underlying ontology, bringing about the analyzed classification. However, in the second place, it has been argued that such ontological assumptions were at odds, in a conceptually fundamental sense, with the tenets of some interpretations of QM.

As a first case, we argued that the ensemble view does not share the assumption that QM is about single, individual systems. On the contrary, this interpretation is based on the idea that quantum mechanics is about \emph{ensembles}, thus adopting a new ontological category not straightforwardly reducible to individuals. In accordance to this, the statistical view poses a $\lambda$ that refers to ensembles, and thereby, $\psi$ should be assessed in virtue of capturing such a new reading of $\lambda$. Importantly, by debunking the assumption that $\lambda$ describes individual quantum systems, we showed that $\psi$ in the statistical view cannot be classified as $\psi-$epistemic. As a consequence of this rationale, we also concluded that more sound arguments and more textual evidence is needed in order to affirm that Einstein supported a $\psi-$epistemic view of the quantum state.

As a second case, we argued that PQM and RQM challenge the idea that $\lambda$ is a complete perspective-free description of quantum systems in terms of intrinsic properties. In contrast, both PQM and RQM put forward a perspectival and relational ontology where $\lambda$ must be regarded as \emph{perspective dependent} and given by relational properties. As a consequence, the $\psi$s should be contrasted with a quantum ontology that admits divergences and overlappings by principled reasons. In the case of a PQM model, we argued that it would be misguided to consider it as $\psi$-epistemic for the interpretation allows the possibility of probability overlapping distributions when a system is described by different perspectives. To be emphatic: a perspectival and a relational ontology do not consider such a possibility as a failure, but as an inherent feature of the quantum world. For these reasons, any PQM model ought to be classified as $\psi$-ontic if it is meant to capture fully a relational and perspective dependent $\lambda$. In the case of a RQM model, the quantum state is indeed regarded as $\psi$-epistemic, but for different reasons as we detailed in Sections \ref{4.2} and \ref{4.3}. Importantly, Harrigan and Spekkens' classification fails to capture such reasons, blurring out the distinctive features of a relational ontology, in general, and of RQM, in particular.

The most general conclusion that can be drawn from our arguments and from the cases we presented is the following. One of the reasons why Harrigan and Spekkens' classification fails to accommodate some interpretations of QM is because it presupposes that $\lambda$ remains fixed across different interpretations: it is taken for granted that in various quantum theories $\psi$ varies, whereas $\lambda$ does not in the precise sense that for the authors $\lambda$ always refers to properties of individual systems inserted into a single, unique perspective. \emph{Contra} these claims, we hold that $\lambda$ can vary across different interpretations. Hence, a more adequate classification should also incorporate divergent views about $\lambda$ as criteria to sort ontological models out. 

This implies that a more adequate classification of quantum models should take into consideration further categories to satisfactorily represent the various features of $\lambda$. Given that we showed that the ontic state of quantum systems does not always refer to a complete description of single, individual systems, a general categorization of quantum models should specify the nature of $\lambda$ as to whether it is meant to completely describe \emph{individual} systems or not. If it does, then a complete description of the properties of a quantum system is a description of properties instantiated by \emph{individual systems}. If it does not, then $\lambda$ is meant to completely describe ensembles. In the second place, it should be also specified whether $\lambda$ is perspective independent or not: in the former case, a complete description of a quantum system would involve intrinsic attributes instantiated by a particular physical system, in the latter case, such a description would take into account relational properties that depend on a some reference system.

Conforming to what we said above, our proposal suggests to introduce new categories to classify ontological models ---i.e.\ the specification of whether is $\lambda-$perspective independent and $\lambda-$individual---in order to show how $\lambda$ may vary across different interpretations. With them, we could reappraise Harrigan and Spekkens' categorizations of the quantum state $\psi$. For instance, their definition of $\psi$-epistemic models relies heavily on $\lambda$ to be perspective independent; by contrast, a far-reaching definition should take this aspect of the ontic state into consideration.

In sum, Harrigan and Spekkens' categories can be made more precise and correct incorporating how different ontological models pose different $\lambda$s. Clearly, these new distinctions we are introducing make the classification of ontological models substantially more complex. However, this is the price to pay in order to comprehend the essential ontological assumptions underlying the various interpretations of QM. In conclusion, a far-reaching and more rigorous classification should take variations of $\lambda$ into account seriously, which entails reappraising the status of $\psi$ in function of such variations. Such a task will be matter of future work.
\vspace{5mm}

\textbf{Acknowledgements:} We warmly thank Olimpia Lombardi, Michael Esfeld and the two anonymous referees for helpful comments on previous versions of this paper.\ AO thanks the Swiss National Science Foundation for financial support (Grant No.\ 105212-175971) and Olga Sarno for helpful discussions on this paper.
\clearpage

\bibliographystyle{apalike}
\bibliography{PhDthesis}
\clearpage
\end{document}